\documentclass[structabstract]{aa}  
\usepackage{graphicx}
\usepackage{txfonts}
\usepackage{natbib}
\bibpunct{(}{)}{;}{a}{}{,} 
\usepackage{color}
\begin{document}
   \title{Detection of a large massive circumstellar disk around a
high-mass young stellar object in the Carina Nebula\thanks{The near-infrared observations for this project were
collected with the HAWK-I
instrument on the VLT at Paranal Observatory, Chile, under
ESO program 60.A-9284(K).
}}

   \author{T.~Preibisch\inst{1} \and T.~Ratzka\inst{1} \and T.~Gehring\inst{1} \and H.~Ohlendorf\inst{1} \and H.~Zinnecker\inst{2,3,4} \and R.~R.~King\inst{5} \and M.~J. McCaughrean\inst{6,5} \and J.~R.~Lewis\inst{7}
          }

   \institute{Universit\"ats-Sternwarte M\"unchen, 
        Ludwig-Maximilians-Universit\"at,
          Scheinerstr.~1, 81679 M\"unchen, Germany
  \and
      Astrophysikalisches Institut Potsdam, An der 
      Sternwarte 16, 14482 Potsdam, Germany
    \and
   Deutsches SOFIA Institut, Universit\"at Stuttgart, Pfaffenwaldring 31,
    70569 Stuttgart, Germany
    \and
NASA-Ames Research Center,
MS 211-3, Moffett Field, CA 94035, USA
   \and
Astrophysics Group, College of Engineering, Mathematics and Physical Sciences, University of Exeter, Exeter EX4 4QL, UK
\and 
European Space Agency, Research \& Scientific Support Department, ESTEC, Postbus 299, 2200 AG Noordwijk, The Netherlands
\and
  Cambridge Astronomical Survey Unit,
  Institute of Astronomy, Madingley Road, Cambridge, CB3 0HA, UK\\
             }

\titlerunning{A large massive circumstellar disk surrounding  a
high-mass young stellar object in the Carina Nebula}
\authorrunning{Preibisch et al.}

   \date{Received  17 January 2011;  accepted 7 April 2011}

 
  \abstract
{The characterization of circumstellar disks around 
young stellar objects can provide important information about the 
process of star formation and
the possible formation of planetary systems.}
{We investigate the spatial structure and the spectral energy
distribution of a newly discovered edge-on circumstellar disk
around an optically invisible young stellar object that is embedded
in a dark cloud in the Carina Nebula.}
{The disk object was serendipitously discovered in our deep
near-IR imaging survey of the Carina Nebula obtained with
HAWK-I at the ESO VLT. Whereas the object 
was detected as an apparently  point-like source in  earlier
infrared observations, only the
superb image quality (FWHM $\approx 0.5''$) of the
HAWK-I data could reveal, for the first time,
 the peculiar morphology of the object.
It consists of a very red point-like central source that
is surrounded by a roughly spherical
nebula, which is intersected by a remarkable dark lane through 
the center.  We construct the spectral energy distribution of the object
from $1\,\mu$m to $870\,\mu$m and perform a detailed radiative transfer modeling
of the spectral energy distribution and the source morphology.
}
{The observed object morphology in the near-IR images clearly
suggests a young stellar object that is 
embedded in an extended, roughly spherical envelope and
surrounded by a large circumstellar disk with a diameter of $\approx 5500$~AU
that is seen nearly edge-on. The radiative transfer modeling shows
that the central object is highly luminous and thus must be a massive
young stellar object, most likely in the range $10\,-15\,M_\odot$.
The circumstellar disk has a mass of about $2\,M_\odot$.
}
{The disk object in Carina 
is one of the most massive young stellar objects for
which a circumstellar disk has been detected so far. 
The size and mass of the disk are very large
compared to the corresponding values found for most other similar objects.
These results support the assumption that $10\,-15\,M_\odot$ stars can form
via accretion from a circumstellar disk.
}
 
   \maketitle
%

\section{Introduction}

During the past dozen years, 
a substantial number of circumstellar accretion disks 
around young stellar objects have been \textit{spatially resolved}
with a number of different techniques such
as direct optical/infrared imaging, (sub-)mm or radio-mapping, 
radio-interferometry, and (more recently) infrared interferometry 
\citep[see, e.g.,][]{Stapelfeldt98,Zinnecker99,Wolf03,Leinert04,Preibisch06,Kraus08,Ratzka09,Gramajo10,Quanz10}.
Spectroscopic observations have also provided important kinematical evidence
of accretion disks \citep[see, e.g.,][]{Chandler95,Bik04,Wheelwright10,Davies10}.
These studies have greatly advanced our understanding of the
physics of young stellar objects (YSOs), how the young stars
gain their final mass, how the circumstellar matter, in which planets may 
be forming,
evolves with time and is finally dispersed
\citep[see][]{Dullemond07}.

So far, nearly all well characterized
circumstellar disks surround low- or intermediate mass 
stellar objects, with less than about $5-8\,M_\odot$.
Only a very limited number of disk detections have been
reported around objects with higher masses
in the range $M \sim 8 - 20\,M_\odot$
\citep[see, e.g.,][]{Chini04,Jiang05,Nielbock07,Kraus10,Davies10}.
However, 
in many of these cases the mass estimates for the central
star are quite uncertain; 
the true mass may in fact be considerably smaller
than initially estimated
\citep[see, e.g.,][]{Sako05}.
Another problem is that the observed flattened structure around 
these more massive young
stellar objects are not necessarily accretion disks,
from which material is directly transferred to the central star; 
they may rather represent a thick torus that could be the remnant
of a flattened envelope \citep{Allen03}.
Considering even higher stellar masses, $M \ga 20\,M_\odot$,
no clear observational evidence for 
accretion disks has yet been found \citep[see, e.g.,][]{Cesaroni07}.
This rarity and lack of convincing disk detections is an 
important aspect in the
long-standing discussion about how massive stars 
 form.
A fundamental problem is that  the
strong radiation pressure resulting from the very high
luminosity of a massive protostars tends to halt the accretion flow
and should thus limit the final stellar mass,
at least in the case of spherical accretion \citep{Kahn74}.
More recent theoretical results
alleviate this problem by mechanisms such as
non-spherical accretion, the ``flash-light effect'',
and ``photon-bubbles'' \citep[see, e.g.,][]{Yorke02,Krumholz09}.
The more accurate treatment of frequency-dependent radiative feedback 
by \citet{Kuiper10}
suggests that stars with masses
well beyond the upper mass limit of spherical accretion can be formed by accretion.
However, due to the limited numerical resolution of these simulations
\citep[e.g., 1.27~AU in the study of][]{Kuiper10}, the reliability of the
numerical results is not fully clear.
As discussed in \citet{Zinnecker07},
there are still strong indications that the formation of
massive stars is {\em not} simply a scaled-up version of the
low-mass star formation process.
Alternative models for the formation of massive stars highlight the
importance of the cluster environment during early star formation
stages \citep[see, e.g.,][]{Bonnell04,Krumholz10},
and processes such as protostellar collisions and
mergers may be possible in the central regions of
very dense clusters \citep[see, e.g.,][]{Bonnell98,Bonnell05,Bally05,Davies06}.

In the context of these open questions, 
every new detection and good characterization
of a circumstellar disk around a massive ($M \ga 8\,M_\odot$)
young stellar object provides an important new mosaic stone
that helps to solve the puzzle of massive star formation.
However, observations of massive young stellar objects are strongly hampered
by the relative rarity of massive stars, their correspondingly
larger average distances (as compared to low-mass stars),
and the extremely short timescales on which massive protostars
evolve. The expected lifetime of a 
disk around a massive young stellar object is very small,
$\la 10^5$~yrs, since
the circumstellar material is quickly 
dispersed by the enormous luminosity and wind power of the
young high-mass stars. 
Furthermore, due to their extreme youth, massive YSOs with disks
are still deeply embedded in their natal
cloud and thus optically invisible. 

\smallskip

The Great Nebula in Carina \citep[NGC 3372; see][for an overview]{SB08}
is a superb location in which to study the 
physics of massive star formation.
At a distance of 2.3\,kpc, it represents the nearest 
southern region with a large population of massive stars,
among them several
of the most massive ($M \sim 100\,M_\odot$)
and luminous stars known in our Galaxy.
Most of these very massive
stars reside in the clusters Trumpler~14 and 16,
which are thought to have ages of a few Myr \citep{Tapia03}.  
Recent sensitive infrared, sub-mm, and radio observations 
showed clearly that the Carina Nebula Complex (CNC)
 is a site of ongoing star formation.
The region contains more than $10^5\,M_\odot$ of gas and dust
 \citep[see][]{SB08,Preibisch11a}.
Several very young stellar objects
\citep{Megeath96,Mottram07} and a spectacular young embedded cluster
\citep[the ``Treasure Chest Cluster''; see][]{Smith05} have been 
found in the molecular clouds, 
 a deep HST H$\alpha$ imaging survey revealed
dozens of jet-driving young stellar objects \citep{Smith10a}, and
Spitzer surveys located numerous embedded protostars  throughout the
Carina complex \citep{Smith10b,Povich11}.
A recent deep wide-field X-ray survey revealed 
at least $\sim 11\,000$ young stars in this area
\citep{Townsley11,Preibisch11b}.
The CNC is 
thus an ideal target in which to  search for very young massive
stars.


%
We have recently used 
the ESO 8~m Very Large Telescope to perform 
the largest and deepest near-infrared survey of the
CNC obtained so far.
In our images, we discovered an optically invisible
infrared source with a 
very interesting morphology, which seems to be a 
young stellar object surrounded by a huge circumstellar
disk seen nearly edge-on. It will be referred to as the ``disk object''
in the following text.
In this paper we first discuss the morphology of the
object and its environment,
and then determine its spectral energy distribution from the
near-infrared to the sub-mm range.
We compare the spectral energy distribution to models of
young stellar objects surrounded by circumstellar disks and envelopes.
Finally, we present the results of our comprehensive and
detailed radiative transfer simulations in which we 
simultaneously modeled the spectral energy distribution and the
observed near-infrared morphology of the object in order to 
determine basic parameters of the central young stellar object and
the physical properties of the surrounding disk and envelope.


\section{Observations}

The near-infrared data presented in this paper were obtained 
in the context of our large survey  of the Carina Nebula
\citep[see][for more details]{Preibisch_HAWKI}
with the instrument HAWK-I \citep[see][]{HAWKI08} at the 
ESO 8m Very Large Telescope. 
HAWK-I is equipped with a
mosaic of four Hawaii 2RG $2048\times 2048$ pixel detectors with
a scale of $0.106''$ per pixel.
All data were processed and calibrated by the Cambridge
Astronomical Survey Unit using pipeline software originally
designed for the analysis of the UKIRT Infrared Deep Sky Survey.
In the final images,
objects as faint as $J \sim 23$~mag, $H \sim 22$~mag,
and $K_{\rm s} \sim 21$~mag are detected at $S/N = 3$. 

The parts of the HAWK-I mosaic in which the disk object is located 
were obtained in the night of 27 January 2008.
The image quality of these specific frames  
is particularly good; we measured the PSF size for several
point-like sources near the disk object and found
FWHM values of $0.37''$ in the $K_s$-band, $0.45''$ in the $H$-band,
and $0.52''$ in the $J$-band.
This excellent image quality (for ground-based seeing-limited observations)
was essential for the detection and
morphological characterization of the disk object.

\smallskip

In order to get more comprehensive information about the
disk object
we also analyzed archival data from the \textit{Spitzer} Space Telescope
(IRAC and MIPS maps)
and optical \textit{Hubble} Space Telescope (HST) images taken with WFPC2 and ACS.

Finally, we also used our own deep sub-mm map of the CNC 
\citep{Preibisch11a} that we recently obtained with the
Large APEX Bolometer Camera LABOCA at the 
APEX telescope.
This map traces the $870\,\mu$m emission at $18''$ angular resolution 
($\cor$ 0.2~pc at the distance of the Carina Nebula).

 \begin{figure}[t]
\includegraphics[width=8.7cm]{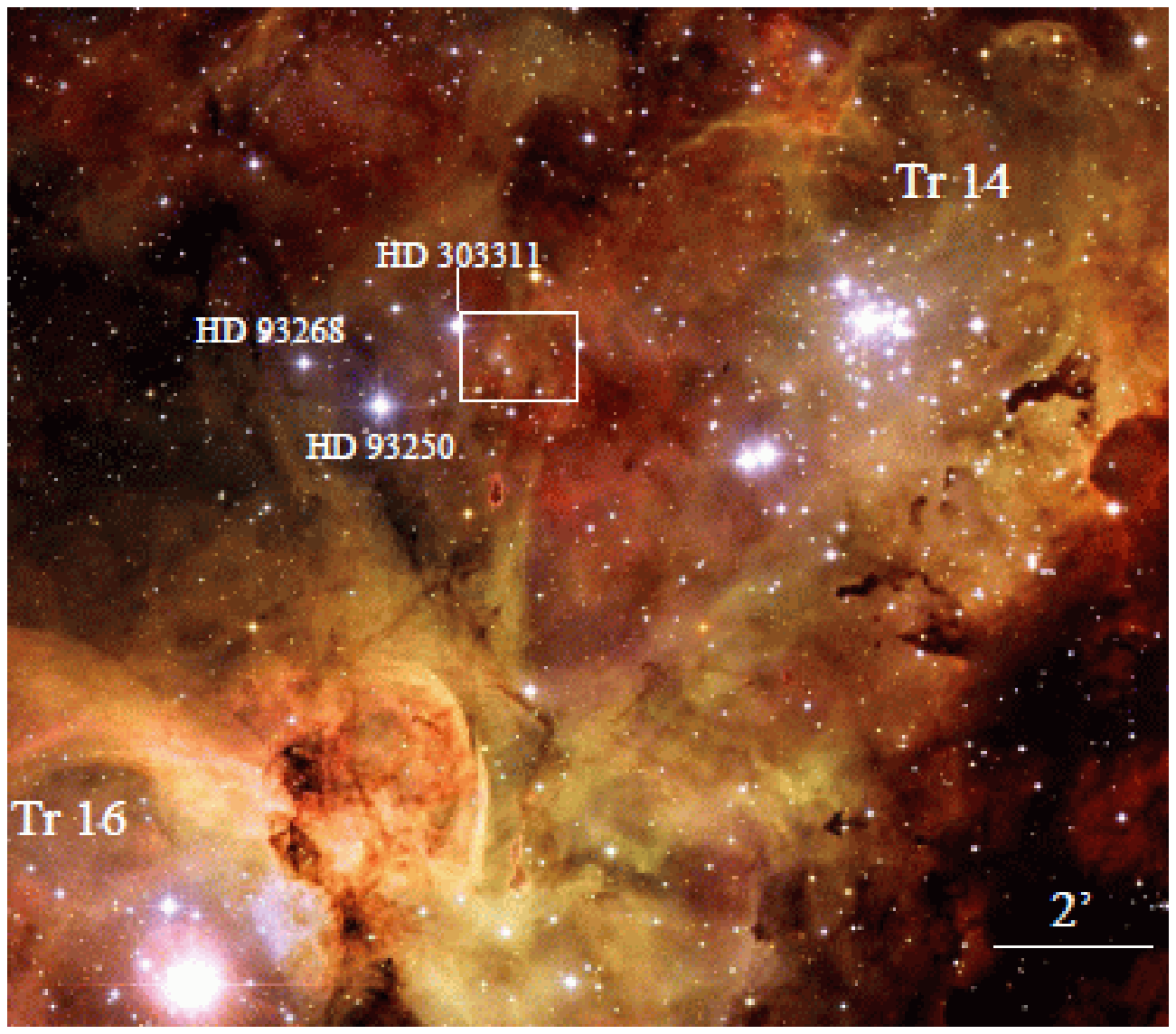}\vspace{1mm}

 \includegraphics[width=8.7cm]{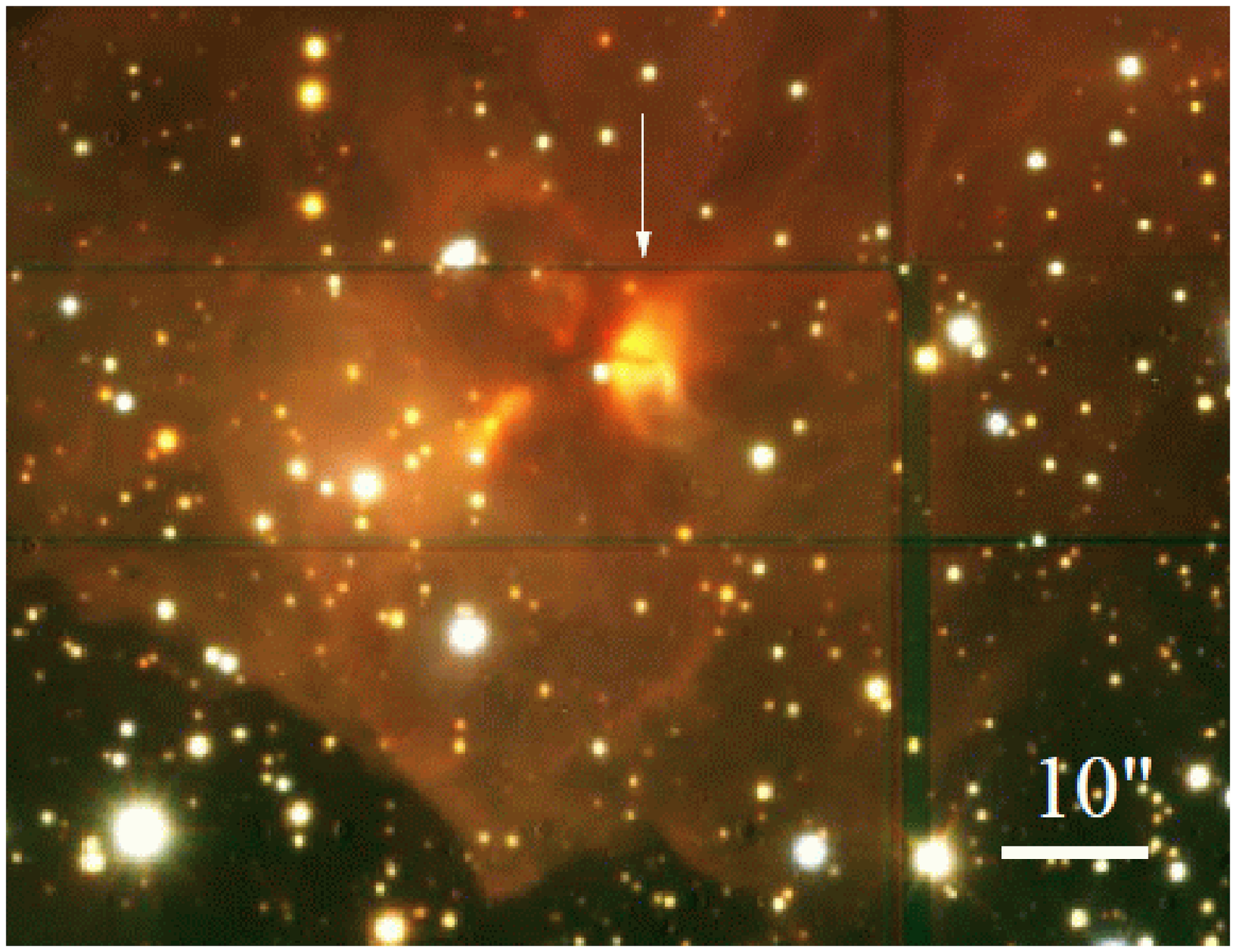}
  \caption{Upper image: Optical image of the central part of the Carina
Nebula, reproduced from the ESO photo release 1031
(http://www.eso.org/public/images/etamosaicnm2/).
North is up and east to the left.
The box shows the
area covered by the near-infrared HAWK-I image below.
\newline
Lower image:
RGB composite image constructed from the $J$- (blue), 
$H$- (green), and $K_s$-band (red) HAWK-I images of the area around the
disk object.
(Note: the horizontal and vertical dark streaks
are artifacts related to the dither pattern and the mosaicing process.)
\label{cr232-fig}}
    \end{figure}

  \begin{figure}
  \includegraphics[width=9cm]{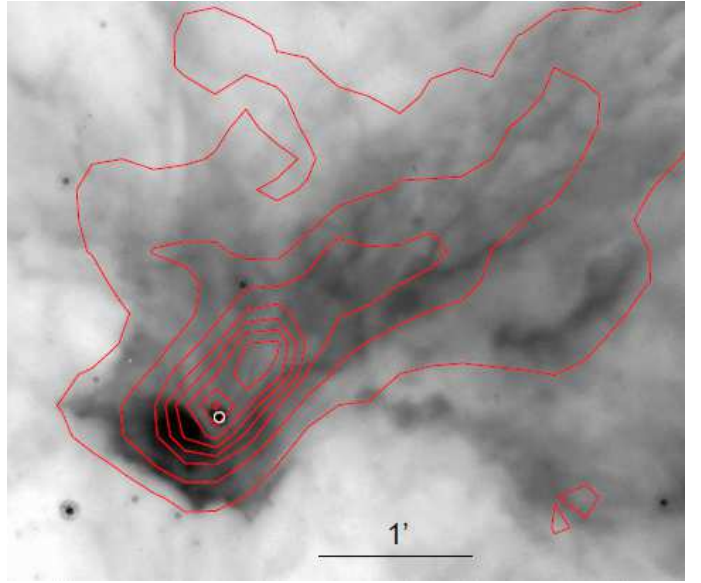}
  \caption{Negative grayscale representation of the $8\,\mu$m Spitzer image
of the cloud in which the disk object is embedded,
with overplotted contours of the $870\,\mu$m LABOCA map.
The field-of-view is $\approx 4.5' \times 3.8'$; north
is up and east to the left. 
The sub-mm contour levels increase 
from 0.06 Jy/beam to 1.5 Jy/beam in equal steps.
The position of the disk object is marked by the
white circle.
\label{8-870-fig}}
    \end{figure}

\section{Discovery of the disk object in the dark cloud near Cr~232}

The disk object was discovered 
in our visual inspection of the HAWK-I images
due to its peculiar morphology. 
It is the brightest member of a group of very red,
deeply embedded objects that are surrounded by extended nebulosity, as
shown in Fig.~\ref{cr232-fig}.
The disk object is located at J2000 coordinates 
$\alpha = 10^{\rm h}\,44^{\rm m}\,31.06^{\rm s}$,
$\delta = -59\degr\,33'\,10.5''$ and 
is seen as 
an extended, roughly spherical nebulosity with a diameter of about $4''$.
The nebulosity is intersected by a prominent dark lane that divides
it into two similarly large hemispheres.
The dark lane can be clearly traced over a distance of $3''$,
corresponding to about 7000~AU.
The global morphology of this object is similar to the famous
`Butterfly Nebula' in Taurus, \citep[a circumstellar disk surrounding a 
T Tauri star, see][]{Wolf03} 
or the  `Silhouette Disk' around a young
star in M~17 \citep{Chini04}, and is discussed in more detail in  
Sect.~\ref{morphology.sec}.

\subsection{The environment of the disk object}

In the optical images of this region (see Fig.~\ref{cr232-fig}) 
the disk object is completely invisible since it is embedded in a
dark cloud in the region of the  clustering Collinder~232.
In order to show the structure of this cloud,
we plot in Fig.~\ref{8-870-fig} the contours of the
 $870\,\mu$m emission (tracing the cold dust in the cloud) on the 
$8\,\mu$m Spitzer image (tracing the
surface of the cloud).
The south-eastern edge of the cloud is very sharp, straight and bright;
its orientation is nearly perpendicular to the direction towards
$\eta$~Car. This suggests that this cloud edge is
an ionization front caused by the strong UV irradiation from
$\eta$~Car and the O-type stars in Tr~16.

The sub-mm emission of this cloud 
shows two peaks. The southern peak 
agrees very well with the position of the disk object, suggesting
that this object dominates the sub-mm emission of this peak.
The second, northern, peak is 
found $19''$ east and $23''$ north of the
disk object. It coincides with the center of the optical dark cloud 
and a corresponding
minimum in the $8\,\mu$m surface brightness (see Fig.~\ref{8-870-fig}). 
No embedded infrared sources
are seen near the center of this second sub-mm peak. 

\subsection{Previous detections of the disk object as an unresolved point-like
 source}
 
  \begin{figure}
  \includegraphics[width=8.5cm]{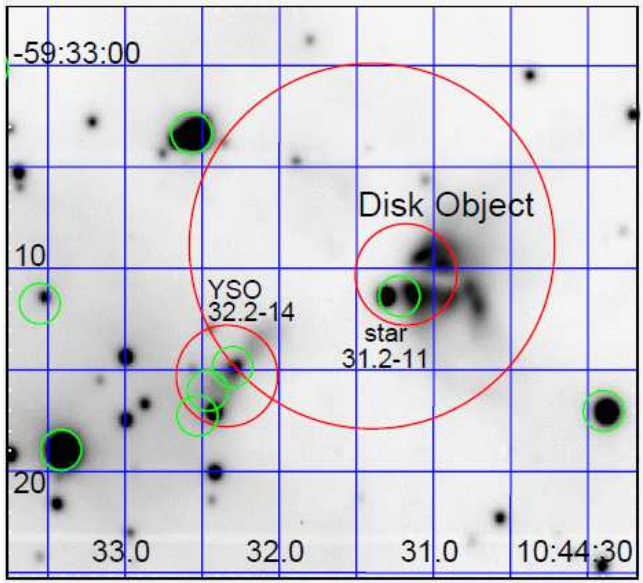}
  \caption{$H$-band HAWK-I image of the area around the disk object.
The small (green) circles  with $2''$ diameter  mark the positions
of the objects listed in the 2MASS point source catalog.
The $5''$ diameter (red) circles mark the locations of the  two
$10.4\,\mu$m sources detected by \citet{Mottram07}, and
the $18''$ diameter (red) circle marks the peak of the
sub-mm emission as seen in our LABOCA map.
A grid of J2000 coordinates is shown.
\label{chart-fig}}
    \end{figure}

Since the disk object is a relatively bright infrared source,
it was detected in many previous infrared observations of this 
region, but the details of its spatial structure always
remained completely unresolved.
Only due to the superior image quality of our HAWK-I data that
provide an angular resolution of $\sim 0.5''$,
its peculiar morphology  could be
resolved.

Figure \ref{chart-fig} shows a small part of the $H$-band HAWK-I image
around the disk source in which the locations of other sources
are marked.
The HAWK-I images show a bright point-like source only
$2.15''$ to the south-east of the disk object.
Its PSF shows no indications of
being extended and it is clearly separated
from the nebulosity surrounding the disk object. We thus believe
that this point source is an unrelated star. 
Based on its position of $\alpha = 10^{\rm h}\,44^{\rm m}\,31.3^{\rm s}$
$\delta = -59\degr\,31'\,11''$ we will refer to it
as {\em star\,31.3-11} in the following.
This star is optically visible and much brighter than 
the disk object in the $J$-band, but less bright than the
disk object in the $K_s$-band. These rather blue colors may suggest
it to be a foreground star.

Inspection of the 2MASS images shows that the disk object and this
nearby point source appear there as an unresolved blend,
listed
as 2MASS-J10443122-5933113 in the point source catalog.
Its position 
is just between the disk object and the south-eastern point source
(as expected).

The disk object is associated with the 
MSX source G287.4779-00.5463,  which appears clearly extended
(FWHM $\sim 40''$) in the MSX images.
The object was also detected 
as a faint diffuse nebulosity in the
thermal-infrared study of the Carina nebula by \citet{Rathborne02},
where it is denoted as source ``N4''.
These authors performed a careful analysis of the
MSX and IRAS data with the aim
to separate the flux from the compact infrared source
from the surrounding diffuse emission 
and derived
the spectral energy distribution of the source from
$8\,\mu$m to $100\,\mu$m (their Fig.~12).
They modeled the object with a two-component blackbody fit to the SED
and derived a total luminosity of $41\,000\,L_\odot$ and a spectral type
of O9--O9.5 for the central object. This result was the first hint that
the disk object is actually a massive young stellar object.

\citet{Mottram07} performed $10.4\,\mu$m mid-infrared observations of G287.4779-00.5463.
Their images were obtained with TIMMI2 at the 3.6~m ESO telescope and
thus provide much better spatial resolution ($\sim 1''$) than the MSX images.
They found two compact mid-infrared sources (these are marked in Fig.~\ref{chart-fig}) 
associated to 
G287.4779-00.5463. The brighter of these mid-infrared sources
coincides perfectly with the disk object,
while the fainter one can be identified with another embedded object with surrounding
diffuse nebulosity seen in our HAWK-I images about $11''$ to the south-east of the
disk object.  Based on its J2000 coordinates of
 $\alpha = 10^{\rm h}\,44^{\rm m}\,32.3^{\rm s}$ $\delta = -59\degr\,33'\,14''$, 
we will refer to this second embedded object as {\em YSO\,32.2-14} in the following.

The angular separation of the disk object and {\em YSO\,32.2-14} is large enough that
they are easily discerned in the \mbox{HAWK-I}, 2MASS, \textit{Spitzer}, and TIMMI2 images.
For MSX and IRAS, however, these two objects cannot be separated
and thus the fluxes derived for wavelengths beyond $10\,\mu$m
are  contaminated by the contribution of the second embedded object.


\section{Morphology and spectral energy distribution 
of the Disk Object \label{morphology.sec}}

  \begin{figure}
  \includegraphics[width=8.75cm]{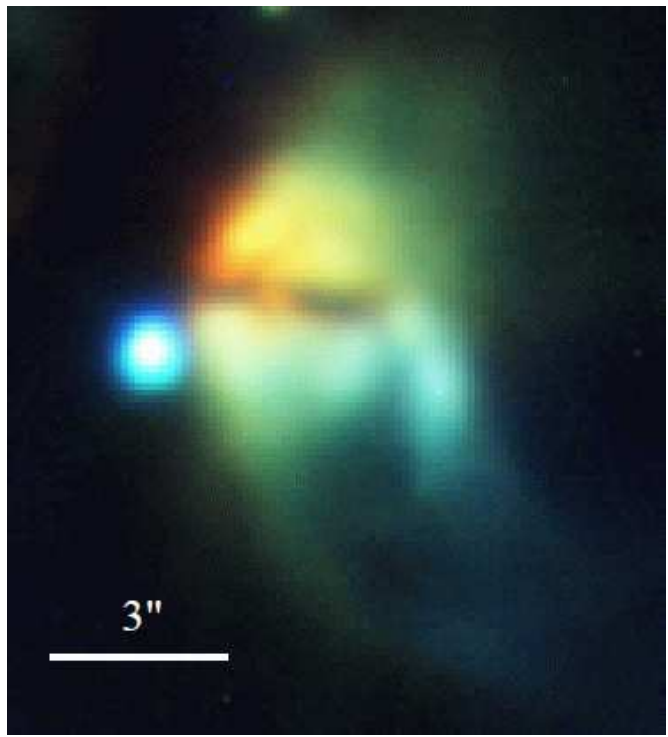}
  \caption{RGB composite image constructed from the  
$K_s$-, $H$-, and $J$-band HAWK-I images.
\label{disk-rgbfig}}
    \end{figure}

\subsection{Near-Infrared Morphology}

A close-up view of the disk object as seen
in our HAWK-I images is shown in 
Fig.~\ref{disk-rgbfig}. The main morphological
structures can be described as follows:

\begin{itemize}
\item The general morphology is that of an
approximately {\bf spherical nebulosity} with
a diameter of about $4''$ (corresponding to 9200~AU) 
and a small degree of elongation, approximately along the
north-south direction.  
\item The nebulosity is divided into a northern and a
southern lobe (hemisphere) by a prominent 
{\bf dark central lane}.
The width of the dark lane is 
$\approx 1.0''$ (2300~AU) in the $J$-band image, $\approx 0.5''$ (1150~AU) in the 
$H$-band image, and
$\approx 0.3''$ (690~AU) in the $K_s$-band image.
\item While the central lane appears completely dark in the $J$-band,
a {\bf faint central object} becomes visible in the  $H$- and $K_s$-band.
\item While the nebulosity has a rather sharp eastern edge,
the western part is much more diffuse. This 
suggests a  {\bf wind-swept morphology} from some agent 
to the east of the object.  We note that there are two known
massive stars in this direction that may be
responsible for this effect:
The O5 star HD~303311,  $48''$ east and $15''$ north of the
disk object (at a projected distance of 0.56~pc), and
the O3 star HD~93250  $105''$ east and $44''$ south
of the disk object (at a projected distance of 1.26~pc).
\item The south western part of the nebulosity
shows some kind of {\bf ``tail''} 
pointing towards the south. This structure
breaks the otherwise high degree of symmetry of the disk object.
As this ``tail'' is the only part of the infrared nebulosity 
that is visible in the optical {\em HST} images,
it might be related to foreground nebulosity.
\end{itemize}


\subsection{Near-Infrared Photometry}

Photometric calibration of the HAWK-I images was achieved via bootstrapping
from aperture photometry measured in the HAWK-I images for the
star 2MASS J10442989-5933170; this star is about $10''$
to the south-west of the disk object, appears well isolated,
and has 2MASS magnitudes of
$J$ = 14.272 $\pm$ 0.04, $H = 13.376 \pm 0.04$, $K_s = 13.015 \pm 0.05$.

For the bright
\mbox{\em star\,31.3-11} near the disk object
we determined magnitudes of 
$J \approx 15.3$, $H \approx 14.6$, and $K_s \approx 14.7$.
The rather bluish color of this star  ($H-K_s = -0.1$)  suggests 
a very low level of extinction and may imply
that this object is located
in front of the dark cloud.
It also
implies that its contribution to the
mid- and far-infrared fluxes for the disk object
will most likely be very small and can
be neglected.

For the disk object, we determined the
following measurements:
The peak surface brightness in the northern\,/\,southern lobes is
17.85\,/\,16.97  mag/sqarcsec in the $J$-band,
15.03\,/\,15.01  mag/sqarcsec  in the $H$-band, and
12.44\,/\,13.13 mag/sqarcsec in the $K_s$-band.
The ratio of the integrated fluxes of the northern and southern lobes
is 
0.62\,:\,1 in the $J$-band, 0.94\,:\,1 in the $H$-band, and
1.47\,:\,1 in the $K_s$-band.
The magnitudes of the central source 
in the $H$- and the $K_s$-band (measured in circular
apertures with 2.5-pixel radii from which
the average of the fluxes measured in two equally large apertures placed 
just to the east and west of the center in the dust lane was subtracted
to account for the local background)
are found to be $K_s \sim 16.1$,
$H \sim 19.6$, $J > 21.8$; note that the errors of these 
measurements are substantial due to the complicated background.
If we assume that the observed emission is actually the central young 
stellar object, the observed color $H-K_s \approx 3.5$
and the assumption that the intrinsic color of the central star should be
$(H-K_s)_0 \leq 0.1$ lead to an extinction estimate of $A_V > 54$ mag.
However, we note that the central patch in the images may also just
be reflected light from the innermost parts of the circumstellar disk;
in that case, the true extinction and optical depth to the central source would
be much higher.

\subsection{Morphology at other wavelengths}

 \begin{figure*} 
  \includegraphics[width=18.3cm]{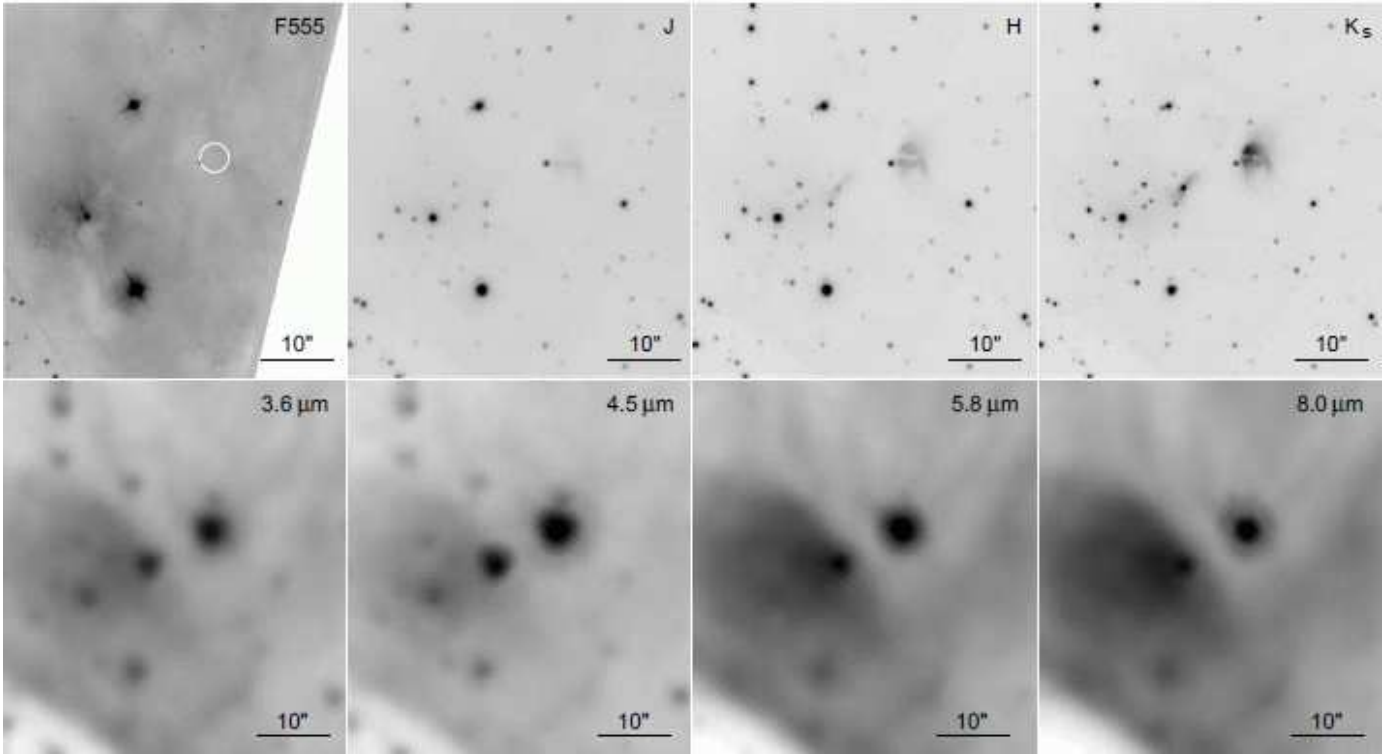}
  \caption{
The disk object and its surroundings seen at different wavelengths.
The upper left panel shows the optical HST image taken through the filter
F555; the position of the (optically invisible) disk object is marked by 
the white circle. The next three panels show the HAWK-I images, and the lower row
shows the \textit{Spitzer} IRAC images.
\label{disk-mw-fig}}
    \end{figure*}

In all optical HST images, \mbox{\em star\,31.3-11} is clearly
visible, but the disk object cannot be seen.
Only the ``tail'' at the south-western edge of the disk object
is visible in the F555W and F850LP image, suggesting that it
may represent reflected light from a {\em foreground}
cloud that is unrelated to the disk object.

In Fig.~\ref{disk-mw-fig} we compare
the Spitzer IRAC images
to our HAWK-I near-infrared images and the HST F555W image.
Although the IRAC images do not have enough angular resolution to
show the structure of the disk object, it
can be seen that the optically invisible disk object
strongly brightens with increasing wavelength.
The same is true for  
\mbox{\em YSO\,32.2-14}.

\subsection{Construction of the spectral energy distribution}

We assembled the spectral energy distribution of the disk object
from the images and literature data discussed above.
For our HAWK-I data we performed aperture photometry using an
aperture diameter of about $5''$, excluding \mbox{\em star\,31.3-11}.
For the Spitzer IRAC data an aperture diameter of $7''$ was used.
Due to the lower angular resolution of the IRAC data, the disk object
and \mbox{\em star\,31.3-11} cannot be separated. However,
as the $K_s$-band flux of \mbox{\em star\,31.3-11} is only $8.8 \times 10^{-4}$~Jy,
i.e.~40 times lower than that of the disk object and since \mbox{\em star\,31.3-11}
shows quite blue colors, we are confident that its contribution to
the emission in the IRAC bands can safely be neglected.
The second embedded source \mbox{\em YSO\,32.2-14}
is clearly separated from the disk object in the IRAC images 
(see Fig.~\ref{disk-mw-fig}) and does not
contaminate our photometry.

Although the Spitzer MIPS maps also have sufficient angular resolution to
clearly separate the disk object from \mbox{\em YSO\,32.2-14},
both objects are  heavily saturated 
and therefore no quantitative photometric analysis is possible.
In order to check the reliability of the
mid- to far-infrared fluxes determined by \citet{Rathborne02} from the
MSX data, we
analyzed and compared the Spitzer MIPS $24\,\mu$m image and the
MSX E-band ($21.3\,\mu$m) image.
The MIPS image shows a bright, roughly circular nebulosity
with a diameter of about $50''$ with three embedded compact, point-like sources.
The brightest of these point-like sources is the disk object, the
second brightest is \mbox{\em YSO\,32.2-14}, and the third one is
the 2MASS source 10443341-5933189 (the bright object near the
lower left corner of Fig.~3). 
In order to determine the $24\,\mu$m flux of the nebula, we used
an aperture with diameter
$54''$ and excluded circular regions with diameters of $11''$ centered
on each of the three point-like sources. Subtracting the large-scale
background level estimated from several nearby regions,
we found a flux of $\approx 46$~Jy for the nebula.
In the MSX E-band image, the cloud is also clearly visible, but
the point-like sources are not resolved.
The total flux, measured again in a $54''$  diameter aperture, is
$\approx 95$~Jy; this is the sum of the emission from the cloud and
the embedded point-like sources.
Subtracting the $\approx 46$~Jy from the nebula as estimated above from the Spitzer map,
leaves a total flux of $\approx 49$~Jy for the sum of the three 
point-like sources. Since the disk object is clearly brighter
in the Spitzer maps than the other two point-like sources,
the flux of 28.4~Jy determined by \citet{Rathborne02} for the dominant
point-like source (i.e.~the disk object) is a very reasonable
value.

We also performed aperture photometry in our LABOCA sub-mm map.
In order to approximately isolate the contribution of the
disk object from the surrounding cloud,
we measured the flux in an aperture with a radius of $18''$ (corresponding to the
FWHM of the beam)  and subtracted the scaled background flux estimated
from the median flux level in an annulus with
1.1 and 1.5 times the aperture radius. This yielded a flux of 2.5~Jy.
As the rather large aperture partly includes 
\mbox{\em YSO\,32.2-14}, this value should be regarded as 
an upper limit
to the sub-mm flux from the disk object.


\begin{table}
\caption{Spectral energy distribution of the disk object.
References: (1)~This work; (2) \citet{Mottram07}; (3) \citet{Rathborne02}.}       
\label{table_fluxes}      
\centering                          
\begin{tabular}{r r r l r}        
\hline\hline                 
$\lambda$ & $F_\nu$ & aperture & Instrument & Reference\\    
 $[\mu\rm m ]$  & [Jy] & diameter & &  \\
\hline                        
1.26 &$0.0013 \pm 0.0003$& $5''$ & HAWK-I &1  \\  
1.62 &$0.0099 \pm 0.0015$& $5''$ & HAWK-I &1  \\  
2.15 &$0.0351 \pm 0.0045$ & $5''$ & HAWK-I &1  \\  
3.6 &$0.113 \pm 0.023$ & $7''$ & Spitzer &1  \\  
4.5 &$0.237 \pm 0.047$ & $7''$ & Spitzer & 1  \\  
5.8 &$0.443 \pm 0.089$ & $7''$ & Spitzer & 1  \\  
8.0 &$0.760 \pm 0.228$ & $7''$ & Spitzer & 1  \\  
10.4 &$0.35 \pm 0.01$ & $1.2''$ & TIMMI2 & 2\\
8.28 & $\approx 5.52$ &  & MSX & 3  \\  
12.13 & $\approx 10.1$ &  & MSX & 3  \\  
14.65 & $\approx 7.3$ &  & MSX & 3  \\  
21.3 & $\approx 28.4$ &  & MSX & 3  \\  
25 & $\sim 50$ &  & IRAS & 3  \\  
60 & $\sim 300$ &  & IRAS & 3  \\  
100 & $\sim 233$ &  & IRAS & 3  \\  
870 & $\la 2.5$ & $36''$ & LABOCA & 1  \\  
\hline                                   
\end{tabular}
\end{table}

\begin{table}
\caption{Spectral energy distribution of
the source \mbox{\em YSO\,32.2-14}}       
\label{table_fluxes_otheryso}      
\centering                          
\begin{tabular}{r r r l r}
\hline\hline                 
$\lambda$ & $F_\nu$ & aperture & Instrument & Reference\\    
 $[\mu\rm m ]$ & [Jy] & diameter  & &  \\
\hline                        
1.26 & $0.00053 \pm 0.00011$& $2''$ & HAWK-I &1  \\
1.62 &$0.0024\pm 0.0036$ & $2''$ & HAWK-I &1  \\
2.15 &$0.0089\pm 0.0013$ & $2''$ & HAWK-I &1  \\
3.6 &$0.053 \pm 0.011$ & $5''$ & Spitzer &1  \\
4.5 &$0.105 \pm 0.021$ & $5''$ & Spitzer & 1  \\
5.8 &$0.219 \pm 0.044$  & $5''$ & Spitzer & 1  \\
8.0 &$0.390 \pm 0.117$  & $5''$ & Spitzer & 1  \\
10.4 &$0.24 \pm 0.01$ & $1.2''$ & TIMMI2 & 2\\
\hline                                   
\end{tabular}
\end{table}

\begin{figure} 
 \includegraphics[width=8.75cm]{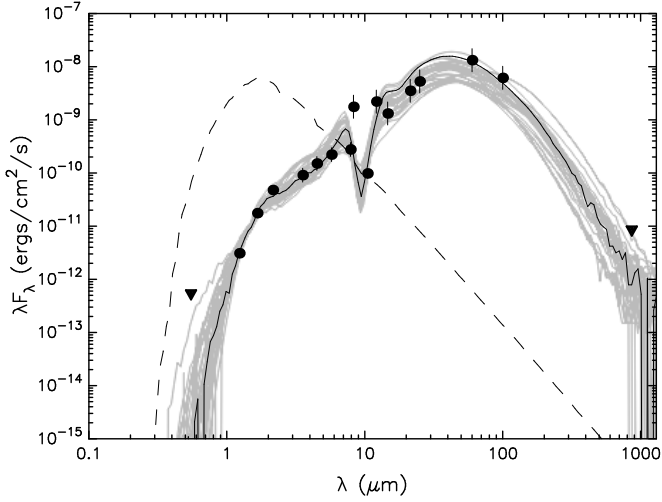}
  \caption{The 32 best fitting Robitaille SED models compared to the
measured fluxes of the disk object. The best fit is shown as a solid line;
the dashed line shows the stellar spectrum of the best fit
model with the effect of the foreground extinction of $A_V = 5.59$~mag.
The grey lines show the other good models as discussed in the text.
\label{robitaille-fig}}
    \end{figure}

The SED of the disk object (see Fig.~\ref{robitaille-fig})
rises steeply from $1.26\,\mu$m
to $8\,\mu$m. The spectral index measured between
$2.15\,\mu$m and $8\,\mu$m is $\alpha = 0.878$; the object thus
qualifies as a ``class I source'', i.e.~a very young object that is 
embedded in large amounts of circumstellar material.

We note again that the
reliability of the fluxes is quite different in the
different wavelength ranges.
The near- and mid-infrared fluxes derived from the HAWK-I and
Spitzer data as well as the TIMMI2 flux are highly reliable and
not contaminated by other infrared sources. The MSX and IRAS fluxes
for the object derived by \citet{Rathborne02} are less reliable
due to the uncertain local background correction and may be contaminated
by emission from \mbox{\em YSO\,32.2-14}.
The LABOCA flux is also uncertain since it may contain significant
amounts of large-scale nebula emission and is therefore regarded as an upper
limit in our modeling described below.

A final caveat comes from the fact that the individual points of the
SED were {\em not} obtained simultaneously.
It is well established that many young stellar objects show considerable
infrared variability by factors of up to $\sim 2$ or even more, 
on timescales between days and decades;
these variations are thought to be
caused by changes in the structure of the inner disks
\citep[e.g.,][]{Flaherty10}.
This implies that flux variations by about a factor of 2 
could be present and change the shape of the observed SED.

\section{Radiative transfer modeling}

\subsection{Modeling of the spectral energy distribution with the Robitaille models}

The pre-computed set of SED models 
provided by \cite{Robitaille07}
are now used as a standard tool to estimate stellar and 
circumstellar parameters of YSOs from the
observed SEDs.
We therefore choose to fit the observed spectral energy distribution
of the disk source with these models as a first step of our analysis.
We note that this kind of analysis ignores the
morphological information we have available from the near-infrared images.
However, since most current studies of YSO
circumstellar structures are based on such SED fits without
spatially resolved image information, this modeling can tell us
what conclusions would be drawn if the direct spatial morphology information
from our HAWK-I images were not available.

Using the SED fluxes as listed in Tab.~\ref{table_fluxes} and 
fixing the distance to 2.3~kpc, the
best fit  ($\chi^2\!/\nu = 1.89$) is found for a model of
a young star surrounded by an envelope with the
following parameters:
$M_\ast =  13.3~M_\odot$,
$L_\ast =  4190~L_\odot$,
$M_{\rm env} = 42~M_\odot$,
$R_{\rm env} = 5.3\times 10^{4}$~AU,
inclination  $= 69.5\degr$, and 
foreground extinction of $A_V = 5.59$~mag.

In the 31 further models with
 $\chi^2\!/\nu - \chi^2\!/\nu\,({\rm best}) < 2$, the parameters are in the following
ranges:
$M_\ast =  8.11 \dots 14.3\,M_\odot$,
$L_\ast =  1320 \dots 4950\,L_\odot$,
$M_{\rm env} = 4.8 \dots 363\,M_\odot$,
inclination  $= 18.2\degr \dots 87.1\degr$, and 
foreground extinction of $A_V = 1.24 \dots 20.9$~mag.
Nineteen of these models have  (in addition to the envelopes)
disks with masses $M_{\rm disk} =8.7\times 10^{-3}\,\dots 4.3\times 10^{-1}\,M_\odot$.

We also performed SED fits excluding the near-infrared fluxes
(that are contaminated by scattered light).
We found that this produces only very marginal changes of the fitting
results.
The best model is the same as the second best model
in the full SED fit.
The parameter ranges for the best 31 models are:
$M_\ast =  8.28 \dots 15.6\,M_\odot$,
$L_\ast =  1550 \dots 6500\,L_\odot$,
$M_{\rm envelope} = 48.2 \dots 251\,M_\odot$,
inclination  $= 18.2\degr \dots 87.1\degr$, and
foreground extinction of $A_V = 0 \dots 35$~mag.

The rather wide ranges for the stellar and circumstellar parameters demonstrate
the large ambiguities of the fits based on the SED only.
If we make use of the information deduced from the observed morphology and
restrict the selection to models with a nearly
edge-on viewing angle ($i > 75\degr$), we find six different models
with the following parameter ranges:
$M_\ast =  8.45 \dots 13.2\,M_\odot$,
$L_\ast =  2140 \dots 4750\,L_\odot$,
$M_{\rm envelope} = 26 \dots 93\,M_\odot$,
$R_{\rm envelope} = 4.4\times 10^{4} \dots 1\times 10^{5}$~AU.
Two of these models also have a circumstellar disk; one with a mass
of $M_{\rm disk} = 0.027\,M_\odot$ and radius $R_{\rm disk} = 122$~AU, 
and the other one with
 $M_{\rm disk} = 0.033\,M_\odot$ and radius $R_{\rm disk} = 8$~AU.

\subsection{2D radiative transfer modeling of the SED and 
morphology}

In the second step of our analysis, we 
performed a more detailed radiative transfer modeling of the object
in which both, the SED {\em and} the near-infrared morphology,
are properly taken into account.
The direct spatial information from the NIR images
can resolve many of the ambiguities of the models
that are based on SED fits only, and provide a much more
physically meaningful model.

We used the 2D radiation transfer code
RADMC \citep{Dullemond04},
which solves the continuum radiative transfer problem in a
passive irradiated
axisymmetric dusty disk and envelope around a central
illuminating star by means of a Monte Carlo algorithm.
After the self-consistent computation of the
dust temperatures throughout the disk, the SED
and images at different wavelengths
are computed with a  ray-tracing procedure.
The density model for the circumstellar disk and envelope
we have used in our modeling
is defined in cylindrical coordinates $(r,z)$
by the following formulae:\newline
The density of the disk is given by
\begin{equation}
 \rho_{\rm disk} = \rho_{\rm disk, 0}\,\left( \frac{r}{r_0} \right)^{-\alpha} \exp\left[-\frac{z^2}{2\,h^2}  \right],
\end{equation}
where $r_0$ is set to the stellar radius and
$h$ is the vertical pressure scale height.
The parameter $h$ can increase with the radial distance from the center, 
resulting in a flaring of the disk:
\begin{equation}
  h(r) = h_{0} \left( \frac{r}{r_0} \right)^{\beta}\;.
  \label{eqn:keplerdiskscaleheight}
\end{equation}

This disk is embedded in a spherical envelope with
 power-law density gradients:
\begin{equation}
 \rho_{\rm env} = \rho_{\rm env, 0}\,\left( \frac{r}{R_{\rm env}} \right)^{\gamma} \;\;{\rm for} \;\; r \leq R_{\rm env}
\end{equation}
and
\begin{equation}
 \rho_{\rm env} = \rho_{\rm env, 0}\,\left( \frac{r}{R_{\rm env}} \right)^{\delta} \;\;{\rm for} \;\; r  > R_{\rm env}\;.
\end{equation}

For our computations, we used a grid size of $R = 5000$~AU,
130 radial grid points,
50 angular grid points 
(with 10 extra angular points to refine the grid near the equator),
a gas-to-dust mass ratio of 100:1, and
$10^5$ photon packets for the Monte Carlo simulations of each model.
For the dust opacities we used the updated model of
\citet{Draine84}\footnote{see
{\tt http://www.astro.princeton.edu/~draine/dust/\newline dust.diel.html}}
with 75\% silicate and 25\% graphite.
We assumed a canonical grain size distribution
$n(a) \propto  a^{-3.5}$ with a minimum grain size
of $0.005\,\mu$m. For the maximum grain
size we used a density dependent value ranging from
$a_{\rm max} = 0.25\,\mu$m   for $\rho < 10^{-17}\,{\rm g/cm^2}$
up to $a_{\rm max} = 10\,\mu$m   for $\rho > 10^{-13}\,{\rm g/cm^2}$.

Our density model is described by eleven free parameters:
(1) Stellar luminosity $L_\ast$; (2) Stellar temperature $T_{\ast}$;
(3) Disk density power law $\alpha$; (4) Disk flaring parameter $\beta$;
(5) Disk vertical scale height $h_0$; (6) Disk density $\rho_{\rm disk, 0}$;
(7) Inner envelope density power law exponent $\gamma$;
(8) Outer envelope density power law exponent $\delta$;
(9) Envelope characteristic radius $R_{\rm env}$;
(10) Envelope density $\rho_{\rm env, 0}$;
(11) Inclination $i$.
Furthermore, as the object is obviously located inside a
dark cloud, we also considered the effects of foreground extinction.

 \begin{figure}
  \includegraphics[width=9cm]{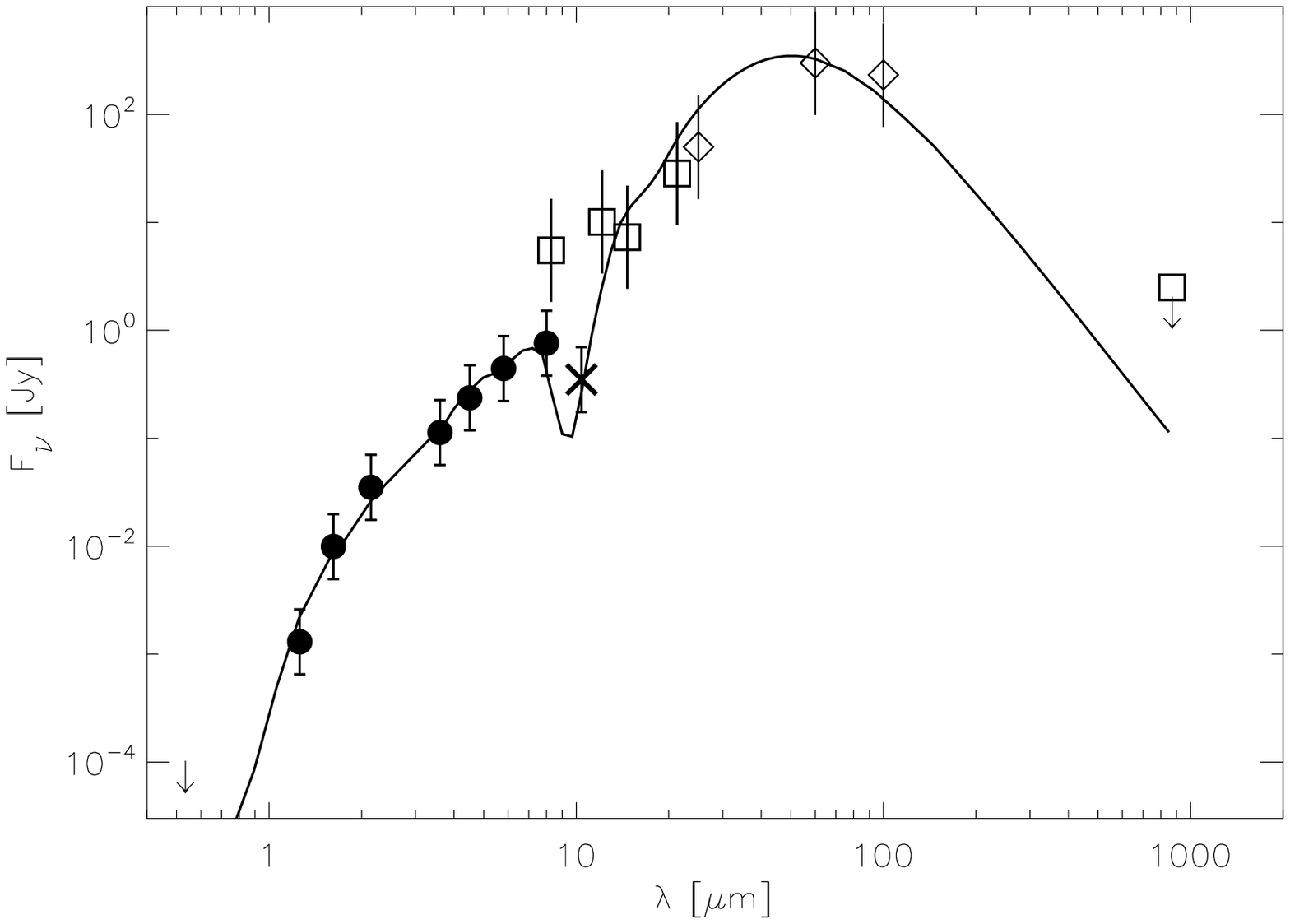}
\parbox{9cm}{
  \includegraphics[width=2.95cm]{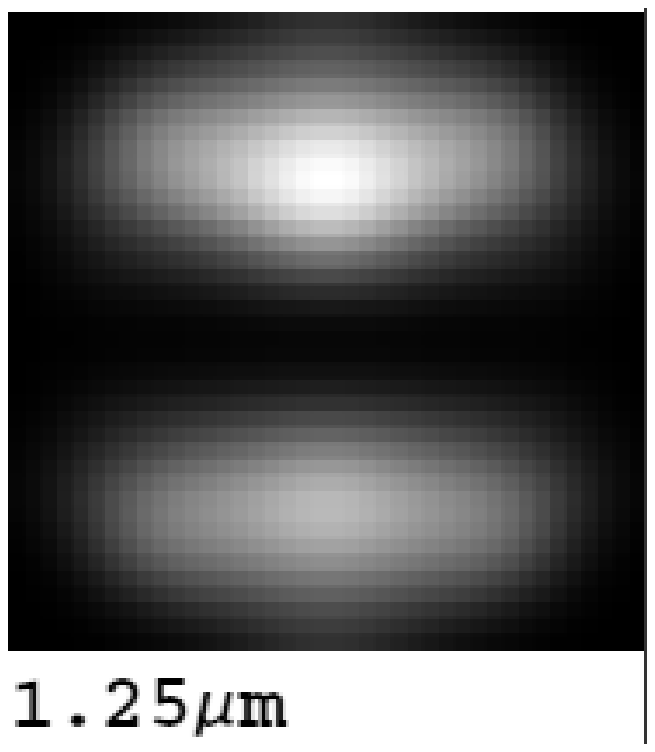}
  \includegraphics[width=2.95cm]{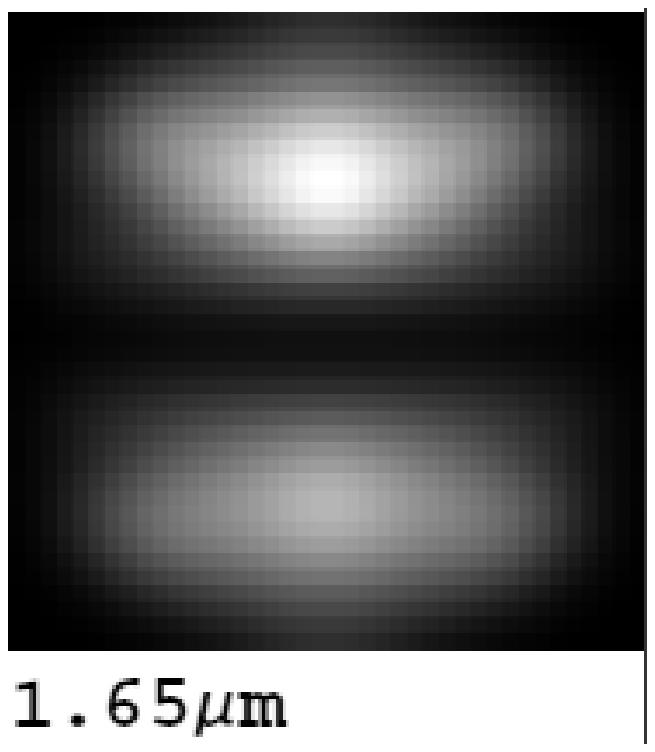}
  \includegraphics[width=2.95cm]{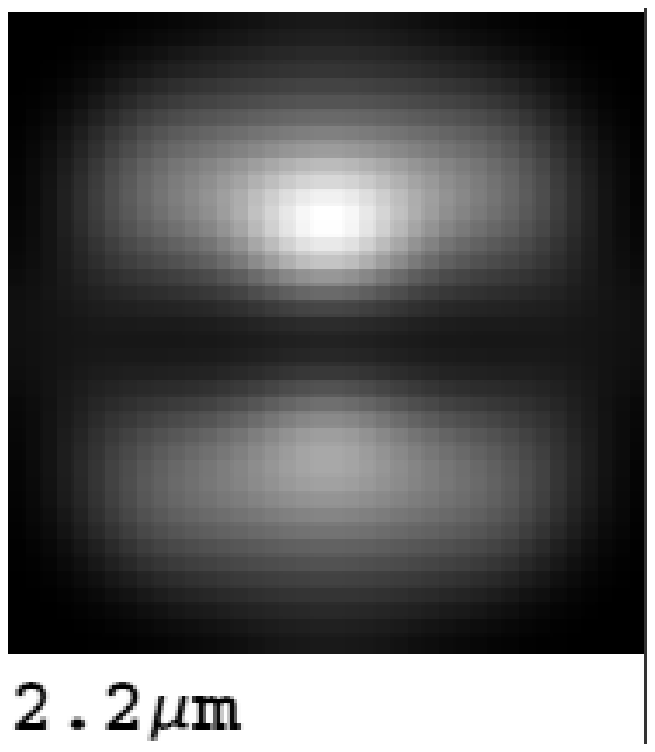}}
  \caption{SED (above) and near-infrared model images (below)
computed with RADMC. The solid dots in the SED plot show the HAWK-I and
\textit{Spitzer} data points, the open squares the MSX data,
the thick \textbf{\sf X} symbol the TIMMI2 measurement, the open diamonds
the IRAS fluxes, and the open square the LABOCA flux. A square-root 
intensity scaling is used for the model images.
\label{radmc-fit}}
    \end{figure}

A general problem of such a radiative transfer modeling
is that the high dimensionality and the complicated topology 
of the parameter space make a search for the
``best model'' very difficult in practice.
Simple scanning of the parameter space is not feasible:
for example, even an extremely coarse
discretization of only 5 different values for each of the
11 parameters
would already require the computation (and evaluation) of about 
50 million different models.

Another problem is how to evaluate the fit-quality of a specific
model. For the comparison of the model SED to the observed
SED, a simple $\chi^2$ analysis is easy to
implement. A quantitative evaluation of the model images,
however, is not so straightforward.
First attempts to compare the individual pixel values in the model images
to those in the observed images were not successful.
Instead, we performed a quantitative assessment of the most
important morphological features in the images.
This was implemented by computing 2D discrete cosine transformations
(DCTs) of the model images and comparing 192 DCT coefficients
per model image
to those derived from the observed images. In this way we made
sure that model images classified as ``good fits''
showed a central dark lane
and a roundish nebulosity above and below this lane.

In the first part of our modeling, we computed 
several thousand models to explore the
main effects of the individual parameters on the
fit quality. Based on these initial results,
we then implemented a  ``genetic algorithm'', in which
small random changes of some of the parameters of good models
are introduced 
in order to find an even  better model. In total, more
than 150\,000 models were computed.

It was very easy to find models that reproduce the
SED very well. However, the images of these models
deviate considerably from the observed images: they generally
produced a 
much thicker dark lane, and often the shape of the 
upper and lower reflection nebulosity was more cone-like
than hemisphere-like (as observed).
On the other hand, we also
found several models that reproduced the near-infrared
morphology quite well; these models, however, could not
reproduce the shape of the observed SED between $2\,\mu$m and 
$\sim 20\,\mu$m.

We finally found several models that
reproduce both, SED and images, in an acceptable way
(although none of these models could {\em simultaneously}
reproduce the SED {\em and} the near-infrared morphology
{\em perfectly}).
One of these ``reasonably good'' models is shown in Fig.~\ref{radmc-fit}.
The model parameters are:
\begin{itemize}
\item Stellar Luminosity: $L_\ast = 12\,000 \, L_\odot$
\item Stellar temperature: $T_{\ast} = 22\,000$~K
\item Disk density power law: $\alpha = 2.25$
\item Disk flaring parameter: $\beta = 1.25$
\item Disk vertical scale height: $h_0/r_0 = 0.2$
\item Total disk mass: $M_{\rm disk} = 1.78\,M_\odot$
\item Inner envelope density power law exponent: $\gamma = 1.3$
\item Outer envelope density power law exponent: $\delta = -4.5$
\item Envelope characteristic radius: $R_{\rm env} = 1500$~AU
\item Total envelope mass: $M_{\rm env} = 0.16\,M_\odot$
\item Inclination: $i = 87\degr$
\end{itemize}
The model assumes a \textit{foreground} (i.e.~interstellar cloud)
extinction of $A_V = 2.5$~mag.
Comparison of the stellar luminosity and temperature
to the pre-main sequence models for massive stars from
\citet{Bernasconi96} suggests a stellar mass of $\approx 12\,M_\odot$.
Such an object should evolve towards a ZAMS star of spectral type $\approx$~B1.

Similarly good models can be found for the following range 
of the main parameters:
$L_\ast \approx 10\,000 \dots 15\,000\, L_\odot$,
$M_{\rm disk} \approx  1.8 \dots 4.7 \,M_\odot$, and
$M_{\rm env} \approx  0.16 \dots 0.5 \,M_\odot$.
The range of values for the vertical disk scale height in the good models is 
$h_0/{r_0} = 0.1 \dots 0.2$; this agrees well with the typical scale heights
found for circumstellar disks around young stars \citep[see][]{Dullemond04}
and clearly shows that the spatial configuration of the circumstellar material
is that of a geometrically thin disk, and \textit{not} a thick torus
(that would be characterized by $h_0/{r_0} > 0.5$).
The luminosity range suggests that the most likely mass of the
central object is in the range $\approx 10 \dots 15\,M_\odot$.
\medskip

One fundamental difference between this model (for SED {\em and}
near-IR image morphology) and the
Robitaille models (for the SED only) is the presence
of a massive and optically thick circumstellar disk.
Such a disk is not required for a good fit to the SED alone,
but its presence is clearly established  from the near-infrared
images.
Another notable difference is seen in the 
values of the stellar luminosity, which
is considerably higher in our RADMC models compared to the
Robitaille SED models.
This difference is related to the so-called ``flashlight effect''
first discussed by \citet{Yorke99}. It results from the highly
non-isotropic distribution of radiative flux that occurs
when a dense, optically thick and flat circumstellar disk is present
around an illuminating central source.
At shorter wavelengths (for which the disk
is optically thick),
the radiation field is strongly concentrated toward the polar directions.
In such a situation, the total observed outgoing radiation flux 
is a strong function of the viewing angle; 
the integrated \textit{observed} flux 
for nearly edge-on inclinations $i \approx 90\degr$ can be
considerably smaller
than the integrated \textit{observed} 
flux measured from a more pole-on view position \citep[see][]{Yorke02}.
In the case of the Carina disk object, the presence of the dark
central lane directly shows that much of the
stellar light is removed from our viewing direction
and scattered into directions perpendicular to the line-of-sight.
Therefore, a high stellar luminosity is required
to reproduce the observed flux seen in our edge-on
viewing direction.
The Robitaille models (that use only the SED as input), on the
other hand, have no such optically thick disk around the star
and thus assume a much more isotropic radiation field.
As a consequence, these models underestimate the luminosity of
the object.

It has to be emphasized that the detection and the 2D
characterization of the disk object was only possible
due to the superb image quality of the HAWK-I data.
In order to check how strongly our modeling results
depend on the angular resolution, we applied Gaussian filters
with different FWHM to the images to see how the observed
morphology changes.
Degrading the intrinsic resolution of the images
($\approx 0.4''$ in the $K_s$-band) to a more typical value
of $0.7''$ is already sufficient to reduce the contrast between the dark
lane and the bright lobes to less than 10\%. 
This implies that under ``average'' seeing conditions the presence of
the dark lane (and thus the nature of the object) would 
no longer be clear.
Since the dark lane is the strongest observational
 constraint for our 2D modeling, the existence of a disk would no
longer be established.

\section{Summary and Conclusions}

Our near-infrared survey of the Carina Nebula
led to the serendipitous discovery of an interesting embedded young stellar object
that is surrounded by a circumstellar disk and a very large envelope.
The presence of the prominent envelope suggests a very early
evolutionary stage of the young stellar object.
The most notable characteristics of this object derived from our
modeling can be summarized as follows:

The central young stellar object has a high
estimated mass of about $12\,M_\odot$; this object is thus
one of the very few known 
cases of massive young stellar objects that are clearly
surrounded by a circumstellar disk.

Another outstanding feature is the
very large physical size of the disk, which  has a
radius of  2700~AU. This is thus much larger than most
disks found around lower-mass young stellar objects (that have typical
radii of a at most a few hundred AU).

Finally,
our modeling suggests a quite high mass of about $2\,M_\odot$ for the
circumstellar disk, and an
additional envelope mass of about $0.2\,M_\odot$.
The total mass of the circumstellar material (within $\la  10\,000$~AU)
seems to be of the order $\sim 2 - 3\,M_\odot$.
This is a remarkably high value for the circumstellar material
around an individual young stellar object. 
Also, the ratio between circumstellar mass and stellar mass is high,
$\sim 20\%$.
This is much higher than typically
found for lower-mass young stellar objects.

Based on these properties, the Carina disk object
clearly represents one of the most interesting circumstellar
disk detections made so far. 
Further observations are planned to study this object in more detail.
An important point is the uncertainty of the current
flux determinations at wavelengths above $\sim 10\,\mu$m, where
the angular resolution of the existing data sets does not
allow an accurate separation of the disk object from neighboring
embedded objects and the surrounding diffuse cloud emission.
Although we have tried to account for these effects,
it is still possible that the SED fluxes
are contaminated by the surrounding nebulosity (as in most other
far-infrared observations of embedded objects). Therefore we cannot exclude the 
possibility that the fluxes from the object are actually somewhat lower, which
could lead to smaller values for the disk mass and the
mass of the YSO.

The radiative transfer modeling demonstrates the essential
value of direct image information.
As mentioned above, the detection and characterization
of the disk was only possible due to the very good angular resolution
of our HAWK-I images and would not have been feasible
under more typical seeing conditions.
This emphasizes (once again) the importance of angular
resolution in the characterization of circumstellar matter around
young stellar objects.
Images with even higher angular resolution (e.g., by adaptive optics
observations) would 
provide very interesting new information about the width and the shape 
of the disk shadow.
In particular, the chromaticity of the
brightness distribution along cuts perpendicular to the disk plane
could provide crucial new constraints for a more detailed
 radiative transfer modeling, and yield
new insight into the physical parameters and dust properties
of the circumstellar material around this massive protostar.

\begin{acknowledgements}
We would like to thank the referee for comments 
that helped to improve the paper.
  We gratefully acknowledge funding of this work by the German
\emph{Deut\-sche For\-schungs\-ge\-mein\-schaft, DFG\/} project
number PR 569/9-1. 
Additional support came from funds from the Munich
Cluster of Excellence: ``Origin and Structure of the Universe''.
This publication makes use of data products from the Two Micron All Sky Survey,
which is a joint project of the University of Massachusetts and the Infrared
Processing and Analysis Center/California Institute of Technology, funded by
the National Aeronautics and Space Administration and the
National Science Foundation.
This work makes use of observations made with the Spitzer Space Telescope, 
which is operated by the Jet Propulsion Laboratory, California Institute of Technology under a contract with NASA.

\end{acknowledgements}

\end{document}